\documentclass[twocolumn]{aastex631}

\usepackage{booktabs}
\usepackage{longtable}
\usepackage{xspace}
\usepackage[T1]{fontenc}

\usepackage{amsmath} 

\newcommand{\Kepler}{\textit{Kepler}\xspace} 
\newcommand{\spitzer}{\textit{Spitzer}\xspace}

\newcommand{\Mstar}{\ensuremath{M_{\star}}\xspace}
\newcommand{\Rstar}{\ensuremath{R_{\star}}\xspace} 
 
\newcommand{\fe}{[Fe/H]\xspace}
\newcommand{\teff}{\ensuremath{T_{\mathrm{eff}}}\xspace}  
\newcommand{\logg}{\ensuremath{\log g}\xspace}

\newcommand{\Me}{\ensuremath{M_{\oplus}}\xspace} 
\renewcommand{\Re}{\ensuremath{R_{\oplus}}\xspace}

\newcommand{\Rsun}{\ensuremath{R_{\odot}}\xspace }
\newcommand{\Msun}{\ensuremath{M_{\odot}}\xspace}

\newcommand{\K}[1]{\ensuremath{K_\mathrm{#1}}\xspace}
\newcommand{\e}[1]{\ensuremath{e_\mathrm{#1}}\xspace}

\usepackage{xstring} 
\usepackage{ifthen} 
\usepackage{xifthen} 

\newcommand{\argperi}[1]{\ensuremath{\ifthenelse{\isempty{#1}}{\omega_P}{\omega_{P,#1}}}\xspace}
\newcommand{\inc}[1]{\ensuremath{\ifthenelse{\isempty{#1}}{i}{i_{#1}}}\xspace}
\newcommand{\ecc}[1]{\ensuremath{\ifthenelse{\isempty{#1}}{e}{e_{#1}}}\xspace}
\newcommand{\per}[1]{\ensuremath{\ifthenelse{\isempty{#1}}{P}{P_{#1}}}\xspace}
\newcommand{\node}[1]{\ensuremath{\ifthenelse{\isempty{#1}}{\Omega}{\Omega_{#1}}}\xspace}
\newcommand{\meananom}[1]{\ensuremath{\ifthenelse{\isempty{#1}}{M}{M_{#1}}}\xspace}
\newcommand{\T}[1]{\ensuremath{\ifthenelse{\isempty{#1}}{T}{T_{#1}}}\xspace}

\newcommand{\ecosw}[1]{\ensuremath{e_{#1} \cos \omega_{#1}}\xspace}
\newcommand{\esinw}[1]{\ensuremath{e_{#1} \sin \omega_{#1}}\xspace}

\newcommand{\zfree}[1]{\ensuremath{\ifthenelse{\isempty{#1}}{z_{\mathrm{free}}^{*}}{z_{\mathrm{free,#1}}}}\xspace}

\newcommand{\chisq}{\ensuremath{\chi^2}\xspace}

\newcommand{\sys}[1]{%
\IfEqCase{#1}{%
{jn-per1}{7.9193}%
{jn-per1_err1}{0.0003}%
{jn-per1_err2}{-0.0002}%
{jn-per1_fmt}{$7.9193^{+0.0003}_{-0.0002}$}%
{jn-tc1}{1980.3838}%
{jn-tc1_err1}{0.0003}%
{jn-tc1_err2}{-0.0003}%
{jn-tc1_fmt}{$1980.4^{+0.0003}_{-0.0003}$}%
{jn-esinw1}{-0.03}%
{jn-esinw1_err1}{0.06}%
{jn-esinw1_err2}{-0.09}%
{jn-esinw1_fmt}{$-0.03^{+0.06}_{-0.09}$}%
{jn-ecosw1}{-0.06}%
{jn-ecosw1_err1}{0.06}%
{jn-ecosw1_err2}{-0.06}%
{jn-ecosw1_fmt}{$-0.06^{+0.06}_{-0.06}$}%
{jn-ecc1}{0.1}%
{jn-ecc1_err1}{0.08}%
{jn-ecc1_err2}{-0.06}%
{jn-ecc1_fmt}{$0.10^{+0.08}_{-0.06}$}%
{jn-omega1}{-115.35}%
{jn-omega1_err1}{251.59}%
{jn-omega1_err2}{-39.62}%
{jn-omega1_fmt}{$-115.35^{+251.59}_{-39.62}$}%
{jn-masse1}{34.9}%
{jn-masse1_err1}{2.0}%
{jn-masse1_err2}{-0.9}%
{jn-masse1_fmt}{$34.9^{+2.0}_{-0.9}$}%
{jn-per2}{11.9064}%
{jn-per2_err1}{0.0012}%
{jn-per2_err2}{-0.0018}%
{jn-per2_fmt}{$11.906^{+0.0012}_{-0.0018}$}%
{jn-tc2}{1984.2738}%
{jn-tc2_err1}{0.0012}%
{jn-tc2_err2}{-0.0012}%
{jn-tc2_fmt}{$1984.3^{+0.0012}_{-0.0012}$}%
{jn-esinw2}{-0.08}%
{jn-esinw2_err1}{0.05}%
{jn-esinw2_err2}{-0.07}%
{jn-esinw2_fmt}{$-0.08^{+0.05}_{-0.07}$}%
{jn-ecosw2}{-0.03}%
{jn-ecosw2_err1}{0.04}%
{jn-ecosw2_err2}{-0.05}%
{jn-ecosw2_fmt}{$-0.03^{+0.04}_{-0.05}$}%
{jn-ecc2}{0.1}%
{jn-ecc2_err1}{0.07}%
{jn-ecc2_err2}{-0.05}%
{jn-ecc2_fmt}{$0.10^{+0.07}_{-0.05}$}%
{jn-omega2}{-108.44}%
{jn-omega2_err1}{30.36}%
{jn-omega2_err2}{-26.8}%
{jn-omega2_fmt}{$-108.44^{+30.36}_{-26.8}$}%
{jn-masse2}{12.5}%
{jn-masse2_err1}{0.5}%
{jn-masse2_err2}{-0.4}%
{jn-masse2_fmt}{$12.5^{+0.5}_{-0.4}$}%
{jn-dwdeg}{14.5}%
{jn-dwdeg_err1}{17.0}%
{jn-dwdeg_err2}{-254.8}%
{jn-dwdeg_fmt}{$14.5^{+17}_{-254.8}$}%
{mstar}{0.88}%
{mstar_err}{0.03}%
{mstar_fmt}{$0.88 \pm 0.03$}%
{rstar}{0.82}%
{rstar_err}{0.03}%
{rstar_fmt}{$0.82 \pm 0.03$}%
{teff}{5322}%
{teff_err}{100}%
{teff_fmt}{$5322 \pm 100$}%
{logg}{4.51}%
{logg_err}{0.08}%
{logg_fmt}{$4.51 \pm 0.08$}%
{z}{0.06}%
{z_err}{0.05}%
{z_fmt}{$0.06 \pm 0.05$}%
{q1}{0.4}%
{q1_err}{0.1}%
{q1_fmt}{$0.4 \pm 0.1$}%
{q2}{0.3}%
{q2_err}{0.2}%
{q2_fmt}{$0.3 \pm 0.2$}%
{u1-tess}{0.4437}%
{u1-tess_err}{0.0018}%
{u1-tess_fmt}{$0.4437 \pm 0.0018$}%
{u2-tess}{0.1452}%
{u2-tess_err}{0.0029}%
{u2-tess_fmt}{$0.1452 \pm 0.0029$}%
{inc1}{91.5}%
{inc1_err}{0.1}%
{inc1_fmt}{$91.5 \pm 0.1$}%
{inc2}{91.1}%
{inc2_err}{0.1}%
{inc2_fmt}{$91.1 \pm 0.1$}%
{ror1}{0.0777}%
{ror1_err}{0.0006}%
{ror1_fmt}{$0.0777 \pm 0.0006$}%
{ror2}{0.0458}%
{ror2_err}{0.0004}%
{ror2_fmt}{$0.0458 \pm 0.0004$}%
{b1}{0.17}%
{b1_err}{0.12}%
{b1_fmt}{$0.17 \pm 0.12$}%
}%
}

\begin{document}

\title{A Decade of Transit-Timing Measurements Confirm Resonance in the K2-19 System}

\author{Paige~M.~Entrican}
\affiliation{Department of Physics \& Astronomy, University of California Los Angeles, Los Angeles, CA 90095, USA}

\author{Erik~A.~Petigura}
\affiliation{Department of Physics \& Astronomy, University of California Los Angeles, Los Angeles, CA 90095, USA}

\author{Antoine~C.~Petit}
\affiliation{Universit\'e C\^ote d'Azur, Observatoire de la C\^ote d'Azur, CNRS, Laboratoire Lagrange, France}

\author{Gregory~J.~Gilbert}
\affiliation{Department of Physics \& Astronomy, University of California Los Angeles, Los Angeles, CA 90095, USA}
\affiliation{Department of Astronomy, California Institute of Technology, Pasadena, CA 91125, USA}

\author{Kento Masuda}
\affiliation{Department of Earth and Space Science, Osaka University, Osaka 560-0043, Japan}

\correspondingauthor{Paige M. Entrican}
\email{pentrican10@g.ucla.edu}

\begin{abstract}
K2-19 is a star, slightly smaller than the Sun, that hosts three transiting planets. Two of these, K2-19 b and c, are between the size of Neptune and Saturn and have orbital periods near a 3:2 commensurability, and exhibit strong transit-timing variations (TTVs). A previous TTV analysis reported moderate eccentricities of $\approx0.20 \pm0.03$ for the two planets, but such high values would imply rapid orbital decay for the innermost planet d. Here, we present an updated analysis that includes eight new transit times from {\em TESS}, which extends the time baseline from three years to a decade, and employ a gradient-aware TTV modeling code. We confirm that the system resides in resonance with a small libration amplitude, but find a broader constraints on eccentricity that range from a few percent up to 0.2. These revised eccentricities alleviate previous concerns regarding rapid tidal circularization and support the long-term dynamical stability of the system. 

\object
\end{abstract}


\section{Introduction}
The existence of mean-motion resonances is one of the most beautiful outcomes of orbital mechanics and has fascinated astronomers as far back as the 18th century \citep{1829mecc.book.....L}. There are several famous and historically significant resonances in the solar system including the 4:2:1 Laplace resonance of Io, Europa, and Ganymede and the 3:2 mean-motion resonance (MMR) that protects Pluto from Neptune. Resonances in exoplanet systems are generally harder to confirm, but a few well-known examples occur in GJ876 \citep{2001astro.ph..8104L}, Kepler-223 \citep{2016Natur.533..509M}, HD~110067 \citep{2024ApJ...968L..12L}, and TRAPPIST-1 \citep{2017NatAs...1E.129L}. Resonances tend to point toward dissipation in a dynamical system and are thus provide information regarding a systems formation. 

The K2-19 system is one of the few exoplanetary systems where two-body MMR has been confirmed. The system hosts three planets discovered in data from NASA's K2 mission Campaign 1. \cite{2015A&A...582A..33A} identified two planets, b and c, that have sizes of 7~\Re and 4~\Re, and orbit at 7.9 and 11.9~days, respectively. Such highly irradiated planets of these sizes are rare. As a point of reference, planets between 4--8~\Re with periods under 10 days occur at a rate of 0.36 planets per 100 stars \citep{2018AJ....155...89P}. Later, \citet{2016ApJ...827...78S} identified a third Earth-size transiting planet, K2-19 d with a period of 2.5 days. 

The proximity of planets b and c to MMR, inspired several groups to obtain follow-up observations with ground- and space-based facilities with goal of detecting TTVs \citep{2015A&A...582A..33A,2015ApJ...815...47N,2015MNRAS.454.4267B,2020AJ....159....2P}. Of the previous analyses, the study by \cite{2020AJ....159....2P} contained the most extensive timing dataset to date with {\em K2} and {\em Spitzer} measurements of planets b and c and ground-based measurements of the larger planet b. The study also included Doppler measurements sufficient to detect planet b but not c; however, the TTVs were far more constraining. The authors also measured eccentricities of the planets $\sim$0.2 which had several key implications for the interpretation of the system. 

The first was that the high eccentricities mean that the `standard' resonant angles associated with the 3:2 MMR circulate rather than librate and led \cite{2020AJ....159....2P} to prematurely conclude the system was not in resonance. However, a follow-on paper by \cite{2020MNRAS.496.3101P} found that the system was indeed in resonance after applying the \cite{1984CeMec..32..307S} transformation to reduce the dynamics to a one-degree-of-freedom integrable Hamiltonian. They found that the angle associated with that dynamical system does, in fact, librate. 

The second was that the high eccentricities of planet b and c would inevitably couple to the inner planet d through the transfer of angular momentum deficit (AMD; \cite{1997A&A...317L..75L}). Planet d is so close to the host star that even modest eccentricities are tidally damped leading to energy dissipation and orbital decay. \cite{2020MNRAS.496.3101P} noted that majority of the orbital configurations published by \cite{2020AJ....159....2P} have orbital decay timescales much less than the age of the system.

Here we present an updated analysis of K2-19, incorporating new timing measurements and analysis techniques that sheds new light on the system's dynamical state and its formation history. We describe the {\em TESS} observations that extend the timing baseline by a factor of $\sim$3 (\S\ref{sec:photometry}) and our analysis with a new gradient-aware TTV code that more robustly explores the high-dimensional parameter space (\S\ref{sec:analysis}). We comment on the resonant state of the system (\S\ref{sec:mmr}), discuss its implications for the system's formation (\S{\ref{sec:formation}}), and offer some concluding thoughts (\S\ref{sec:conclusions}).


\section{Transit Times}
\label{sec:photometry}

\subsection{Literature transit times}
\label{sec:obs-lit}

We compiled 13 previously published transit times for K2-19b and K2-19c from the literature. These included K2 timing measurements from \citet{2015ApJ...815...47N} as well as ground-based measurements from the FLWO, TRAPPIST, and MuSCAT telescopes from the same publication. To this, we added four transits from \spitzer \cite{2020AJ....159....2P} as well as a LCO timing measurement from the same paper. We include these literature times in Appendix~\ref{sec:transit_times}.

\subsection{TESS}
\label{sec:obs-tess}

\begin{figure*}
\centering
\includegraphics[width=0.8\textwidth]{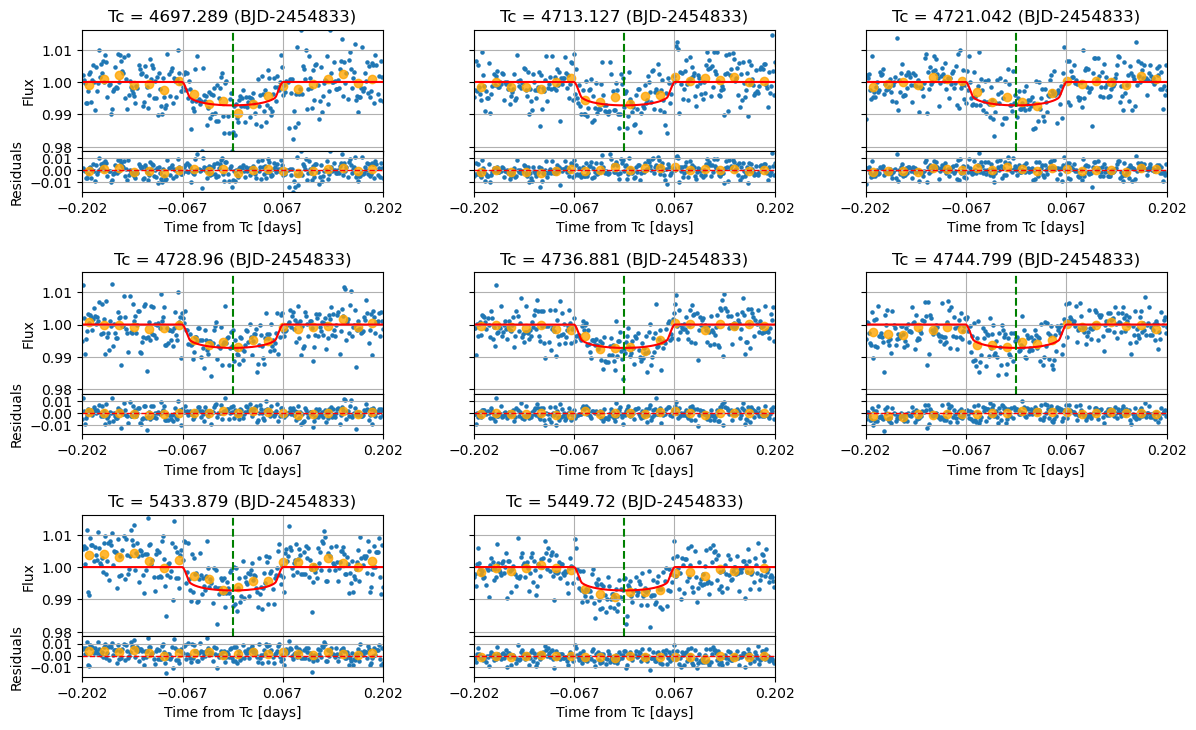}
\caption{Individual transit fits for {\em TESS} transit detections of K2-19~b. The green dashed lines mark mid-transit times, $t_0$. \label{fig:transit-validation}}
\end{figure*}

{\em TESS} observed K2-19 during the following date ranges: from 2021-11-06 to 2022-12-30 (sectors 45 and 46) and from 2023-11-11 to 2023-12-07 (sector 72). We retrieved the SPOC photometry using the Lightkurve Python package \citep{2018ascl.soft12013L} which queries the Mikulski Archive for Space Telescopes (MAST). We extracted 48~hour segments around each of the forecasted transit times from \citet{2020AJ....159....2P}. We show these segments in Figures~\ref{fig:transit-validation}; the transits of planet b are visible by eye, but not planet c.

To measure the transit times, we generated a Mandel-Agol transit model \citep{2002ApJ...580L.171M} using the Batman Python package \citep{2015PASP..127.1161K} and fixed the period, radius ratio, duration, impact parameter, and limb darkening parameters to the tabulated values from the NASA’s Exoplanet Archive \citep{2013PASP..125..989A}. We varied the remaining parameter, the transit mid-point, and evaluated $\chi^2$ over a fine grid of transit times, similar to the approach of \citet{2013ApJS..208...16M} and \citet{2016ApJS..225....9H}. Our reported transit times correspond to the $\chi^2$ minima, and uncertainties were determined by noting the times where $\chisq - \chi^2_\mathrm{min} = 1$ \citep{1992nrfa.book.....P}. The individual transit fits for planet b are shown in Figure \ref{fig:transit-validation}.

Given that the transits of planet c are roughly one quarter the depth of b, it is not surprising that they are lost in the noise. To guide the eye, we generated a preliminary timing model for planets b and c following the approach described in the following section. We plot transits at the expected locations and find that the data is simply too noisy to permit robust timing measurements. We therefore only augment the existing dataset with planet b transits. We considered the possibility that even sub-significant transits of planet c could influence the planet b times. Although some predicted transits of planet c fall relatively close in time to those of planet b, they do not overlap as shown in Appendix~\ref{sec:overlapping-transit}.

\begin{figure*}
\centering
\includegraphics[width=0.7\textwidth]{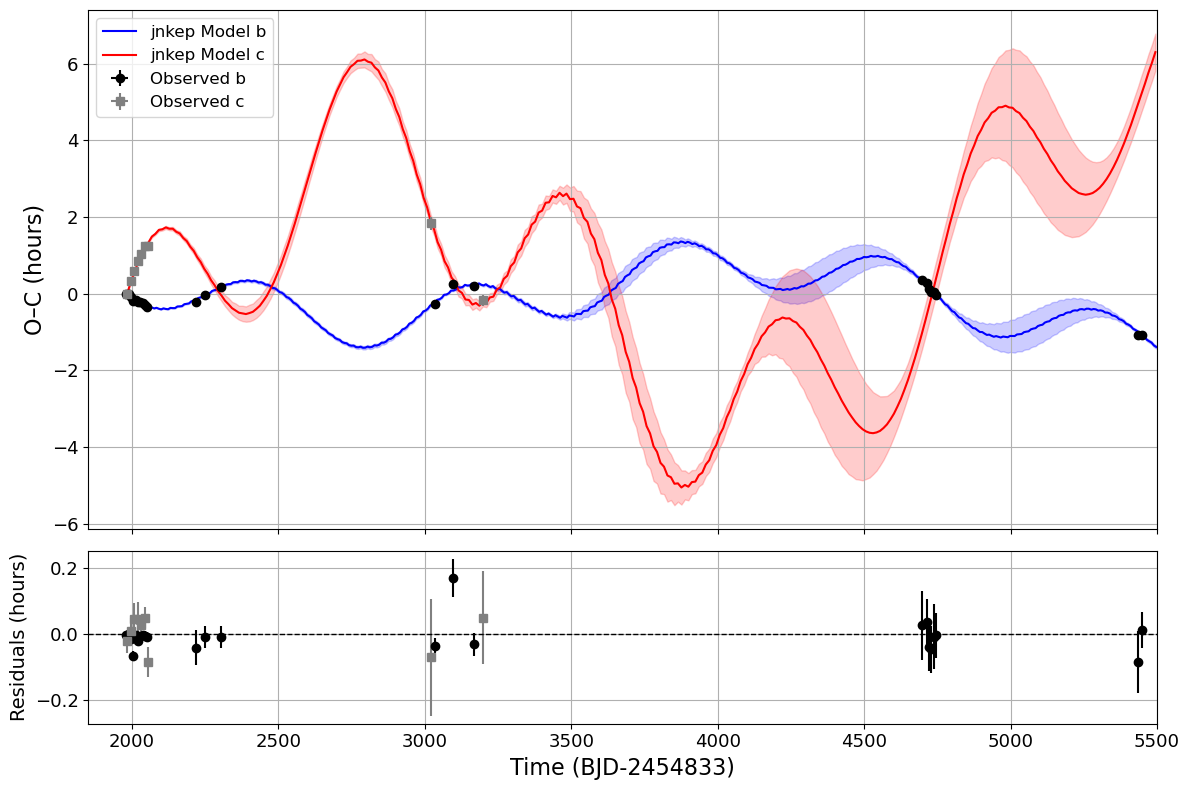}
\caption{The black points represent observed transit times for planet b and the grey points represent observed transit times for planet c. The solid lines represent the model derived in this work, including the TESS transit data. \label{fig:jnkep-model}}
\end{figure*}

Appendix~\ref{sec:transit_times} presents our complete set of timing measurements. Figure~\ref{fig:jnkep-model} shows the transit-timing variations themselves in a `observed minus calculated' or `O-C' diagram. The measured timing variations are large, as much as $\sim$1~hr for planet b and $\sim$5~hrs for planet c.

\begin{figure*}
\centering
\includegraphics[width=0.8\textwidth]{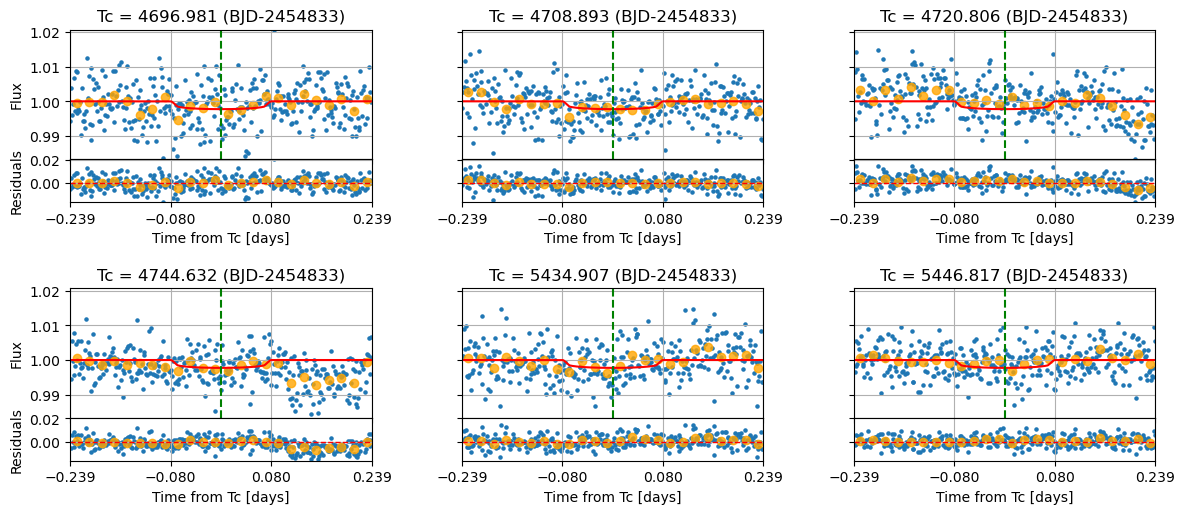}
\caption{Same as Figure~\ref{fig:transit-validation} but for K2-19~c. The green dashed lines show the {\em predicted} transit times based on the MAP model described in \S\ref{sec:obs-tess}. The transits themselves are lost in the noise, and we do not attempt to fit them. \label{fig:c-prediction}}
\end{figure*}

\section{TTV Modeling}

\begin{figure*}
\centering
\includegraphics[width=0.7\textwidth]{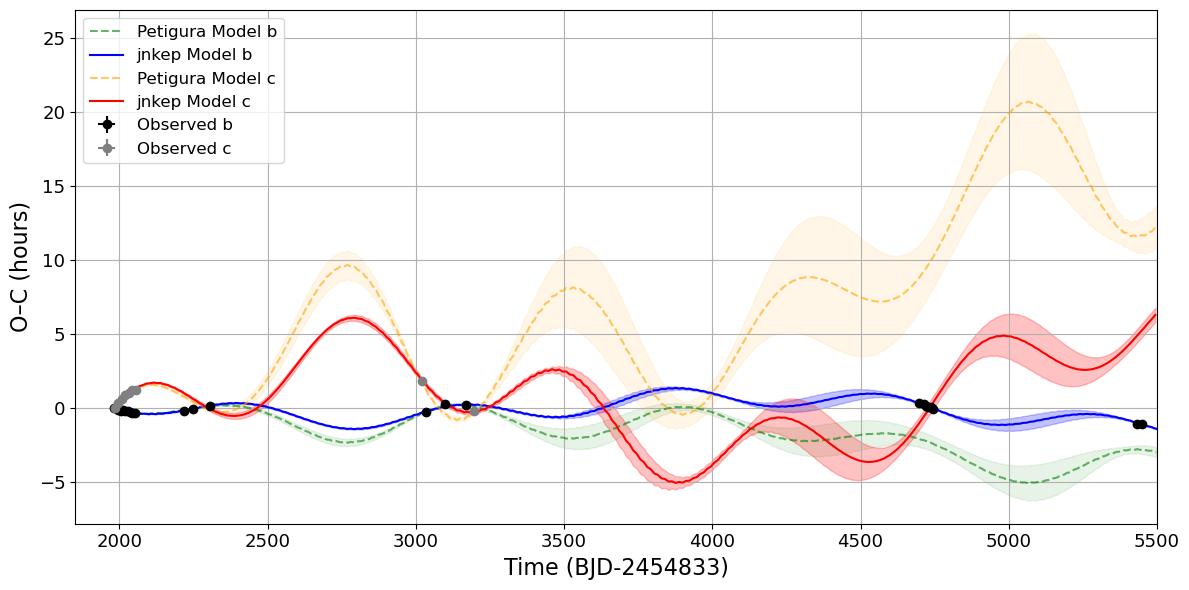}
\caption{Black and gray points are observed TTVs for planet b and c respectively. The dashed lines show the \cite{2020AJ....159....2P} predicted times conditioned on data up to $t \approx 3200$~days; the solid lines show our model that includes all the times shown. The models agree up to $t \approx 3200$~days, but diverge beyond that point.\label{fig:model-compare}}
\end{figure*}

\label{sec:analysis}

\subsection{TTV analysis}
\label{sec:jnkep-model}

We modeled the TTVs using {\em jnkepler} \citep{2024AJ....168..294M}, a differentiable $N$-body integrator designed for multi-planet systems and written in Python. Given a system's initial dynamical state---orbital period $P$, time of inferior conjunction $T_c$, eccentricity $e$, argument of periapsis $\omega$, inclination $i$, longitude of ascending node $\Omega$, and mass $M$ for each planet, along with stellar mass $M_\star$---{\em jnkepler} performs a forward symplectic integration \citep{1991AJ....102.1528W} to determine the state of the system at later times and compute subsequent transit times along the way. A key feature of {\em jnkepler} is its ability to compute the derivatives of transit times with respect to the input parameters using the JAX autodiff library \citep{jax2018github}. This enables efficient sampling of TTV posteriors that often exhibit thin, curving degeneracies. 

As an initial step, we performed a Maximum A Posteriori (MAP) fit to the transit times, assuming Gaussian measurement uncertainties. The RMS of the residuals to the MAP model was 2.9 and 3.4~min for planets b and c respectively; the reduced chi-squared of the MAP model was $\chi^2_\nu = 1.66$.

We explored the range of credible models using Hamiltonian Monte Carlo (HMC) and the No U-turn Sampler (NUTS) \citep{2011arXiv1111.4246H, 1987PhLB..195..216D, 2017arXiv170102434B} as implemented in NumPyro \citep{2018arXiv181009538B, 2019arXiv191211554P}. We initialized four independent chains from the MAP model and evolved them for 2000 steps, discarding the first 1000 as burn-in. The effective sample size ($n_\mathrm{eff}$), aggregated across all chains, exceeded 500 for every parameter. The Gelman–Rubin statistic ($\hat{R}$; \cite{1992StaSc...7..457G}), a split-chain convergence diagnostic that compares intra- and inter-chain variation, was 1.02 or lower, providing evidence of convergence.

Figure \ref{fig:jnkep-model} shows transit times drawn from our set of credible models in addition to the MAP fit. Most of the data falls within 1-2$\sigma$ of the model, indicating that the model is a good fit to the observed transit times. While the overall model provides a good fit to the observed transit times, we identify one notable outlier near $\mathrm{BJD}-2454833 \approx 3200$, which lies significantly above the model. This point is a ground-based measurement, and underestimated timing uncertainties or unmodeled systematics are a concern. We provide constraints on the complete set of parameters Table~\ref{tab:params} and a corner plot in Appendix~\ref{sec:model-post}.

\renewcommand{\arraystretch}{0.82}
\begin{table}
\begin{center}
\caption{K2-19b \& c System Parameters}
\label{tab:params}
\begin{tabular}{lrl}
\hline
\hline
Parameter              & Value   & Notes \\
\hline
\multicolumn{2}{l}{{\bf Stellar}} \\
\teff (K)                 & \sys{teff_fmt}        & A \\
\logg (dex)               & \sys{logg_fmt}        & A \\
\fe (dex)                 & \sys{z_fmt}           & A \\
\Mstar (\Msun)            & \sys{mstar_fmt}       & A \\
\Rstar (\Rsun)            & \sys{rstar_fmt}       & A \\
$u_{1,TESS}$              & \sys{u1-tess_fmt}     & B \\
$u_{2,TESS}$              & \sys{u2-tess_fmt}     & B \\
\multicolumn{2}{l}{{\bf Planetary}} \\
$P_b$ (days)              & \sys{jn-per1_fmt}        & C \\
$T_{c,b}$ (BJD$-$2454833)  & \sys{jn-tc1_fmt}         & C \\
$\ecosw{b}$              & \sys{jn-ecosw1_fmt}     & C \\
$\esinw{b}$              & \sys{jn-esinw1_fmt}     & C \\
$\inc{b}$ (deg)           & 90 (fixed)        & C \\
$\Omega_b$ (deg)          & 0 (fixed)                & C \\
$M_{p,b}$ (\Me)           & \sys{jn-masse1_fmt}  & C \\
$P_c$ (days)              & \sys{jn-per2_fmt}        & C \\
$T_{c,c}$ (BJD$-$2454833)  & \sys{jn-tc2_fmt}         & C \\
$\ecosw{c}$              & \sys{jn-ecosw2_fmt}     & C \\
$\esinw{c}$              & \sys{jn-esinw2_fmt}     & C \\
$\inc{c}$ (deg)           & 90 (fixed)       & C \\
$\Omega_c$ (deg)          & 0 (fixed)      & C \\
$M_{p,c}$ (\Me)           & \sys{jn-masse2_fmt}      & C \\
\multicolumn{2}{l}{{\bf Derived Parameters}} \\
$\ecc{b}$                 & \sys{jn-ecc1_fmt}          & D \\
$\ecc{c}$                 & \sys{jn-ecc2_fmt}          & D \\

\hline
\\[-6ex]
\end{tabular}
\end{center}
\tablecomments{A: \cite{2020AJ....159....2P}. B: Calculated using \citet{Parviainen2015}, accessing the spectrum library described by \citet{Husser2013}. C: Input parameters into Jnkepler $N$-body model, see Section~\ref{sec:jnkep-model}. D: Derived from the posterior samples of C. }
\end{table}

\subsection{Comparison with Petigura et al. (2020)}

Our analysis improves the TTV baseline and phase coverage compared to  \citet{2020AJ....159....2P}. Figure \ref{fig:model-compare} compares the model shown in Figure \ref{fig:jnkep-model} to the model from \cite{2020AJ....159....2P}, which only included times up to $\mathrm{BJD}-2454833 = 3200$. Both models agree over the time span of the original data, but the models diverge at later epochs. This divergence stems from the multi-harmonic nature of the TTV, which in resonant systems arises from variations in mean longitude $\lambda$ and the quantity $e \sin \omega$ where $\omega$ is the argument of pericenter, each with their own periodicity \citep{2016ApJ...823...72N}. Our updated fits clearly show two strong sinusoidal components with periods of $\sim$700 and $\sim$2600~days. The earlier work could only sample the shorter periodicity, leading to divergence at later times. 

The credible models reported here should, in principle, be a subset of the models reported in \cite{2020AJ....159....2P} since they used a subset of the dataset here. However, we find that, over the times plotted in Figure~\ref{fig:model-compare}, the models differ by $\sim$3.5$\sigma$. We hypothesize that this discrepancy results mainly from the earlier analysis's use of a Differential Evolution Markov Chain Monte Carlo scheme (DE-MCMC). Unlike {\em jnkepler}, DE-MCMC does not use gradient information and thus struggles to explore thin, curving degeneracies often present in TTV posteriors. In addition, we note that the outlier timing measurement discussed earlier may have had a greater pull on the earlier analysis.

\begin{figure*}[t!]
\centering
\includegraphics[width=0.4\linewidth]{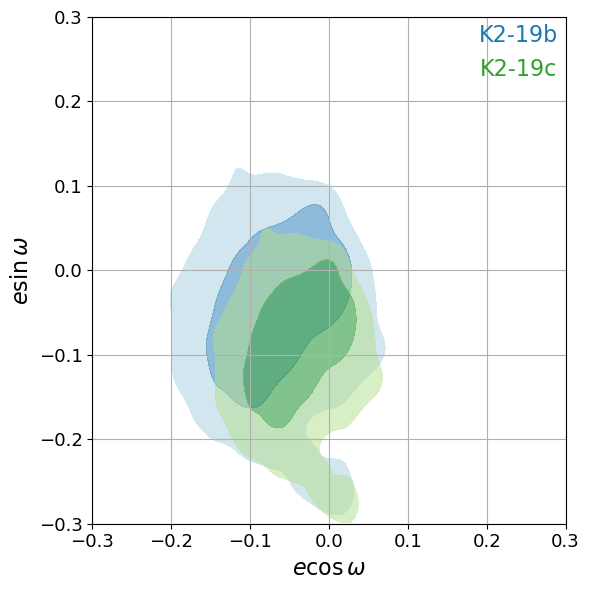}
\includegraphics[width=0.4\linewidth]{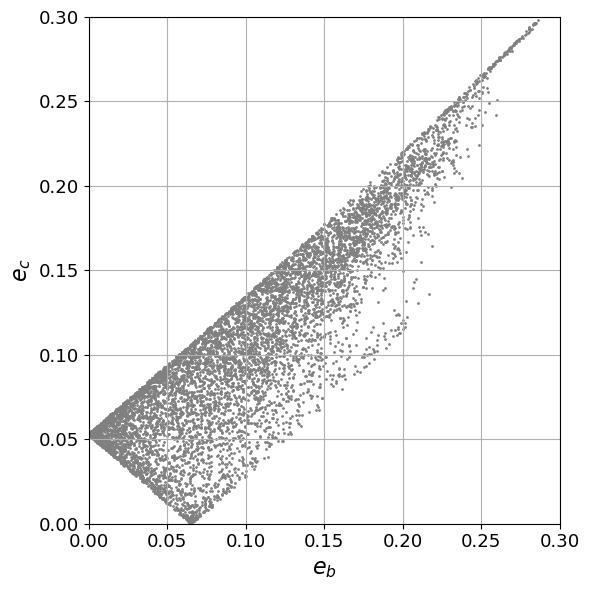}
\caption{Constraints on the eccentricities of K2-19~b and c. Left panel: 2D joint posterior of $e \cos\omega$ and $e \sin\omega$ for K2-19b (blue) and K2-19c (green). The contours represent the $1\sigma$ and $2\sigma$ levels. Right panel: draws from our credible models showing $e_b$ against $e_c$. While either planet may have zero eccentricity, both may not. This arises from our tight constraint on the dynamical variable $Y_1$ described in \S\ref{sec:mmr}\label{fig:ecc-post}}
\end{figure*}

We measured masses of \sys{jn-masse1_fmt}\ and \sys{jn-masse2_fmt}\ \Me, respectively, which are consistent with \cite{2020AJ....159....2P} at $2\sigma$. However, our eccentricity constraints are much broader. Figure~\ref{fig:ecc-post} shows our constraints on the magnitude and orientation of each planet's eccentricity vector. Either planet may have eccentricities as high as $\sim$0.2 or as small as zero. We can, however, confidently state that both eccentricities may not be zero and that $e_c \geq  0.05 - e_b / 1.22$, for reasons that will become clear in the following section. Our eccentricity measurements are in tension with \cite{2020AJ....159....2P}, who measured moderate eccentricities of $e_b=0.20 \pm 0.03$ and $e_c=0.21\pm 0.03$. While the eccentricity measurements are consistent, our constraints are much broader despite our more extensive dataset. Again, we attribute this to the incomplete sampling of the DE-MCMC used in \cite{2020AJ....159....2P}.


\section{Mean-Motion Resonance}
\label{sec:mmr}

We investigated the resonant state of the K2-19 b-c pair using the formalism developed by \citet{1984CeMec..32..307S}, \citet{2013A&A...556A..28B}, and \citet{2016ApJ...823...72N}. Our goal was to assess whether the planets are participating in a 3:2 mean-motion resonance (MMR). 

The Hamiltonian that describes two planet dynamics in (or near) first order resonance is
\begin{equation}
\mathcal{H} = \mathcal{H}_{\text{kep}} + \mathcal{H}_{\text{res}} + \mathcal{O}(e^2, i^2),
\label{eqn:hamiltonian-full}
\end{equation}
where
\begin{equation}
\mathcal{H}_{\text{kep}} = -\frac{G M m_1}{2a_1} - \frac{G M m_2}{2a_2},
\end{equation}
and
\begin{multline}
\mathcal{H}_{\text{res}} = -\frac{G m_1 m_2}{a_2} \Big[
f^{(1)}_{\text{res}} \, e_1 \cos(k\lambda_2 - (k - 1)\lambda_1 - \varpi_1) \\
+ f^{(2)}_{\text{res}} \, e_2 \cos(k\lambda_2 - (k - 1)\lambda_1 - \varpi_2)
\Big].
\end{multline}
Here, $f$ is a constant, $k$ is an integer specifying the resonance, $\lambda$ is the mean longitude, and $\varpi$ is the logitude of percenter. The full Hamiltonian is a function of eight variables, but may be reduced through a series of Canonical transformations (see \citealt{1984CeMec..32..307S,1986CeMec..38..335H,1986CeMec..38..175W}) to the following integrable form: 
\begin{equation}
    \mathcal{H} = \hat{\delta}(\Omega + \Psi_1 + \Psi_2) - (\Omega + \Psi_1 + \Psi_2)^2 - \sqrt{2\Psi_1} \cos(\psi_1)
\end{equation}
Here, $\Psi_1$ and $\psi_1$ are the action and angle that describe the dynamical evolution of the system. The other parameters are constants of motion. 

\begin{figure*}
\centering
\includegraphics[width=.40\textwidth]{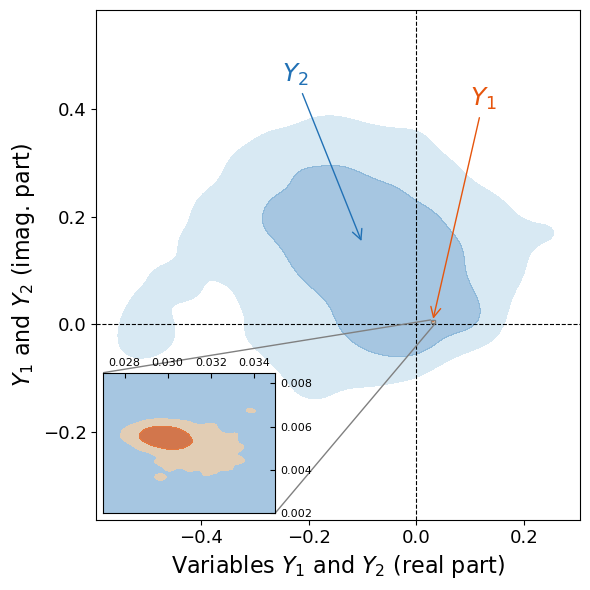}
\includegraphics[width=.54\textwidth]{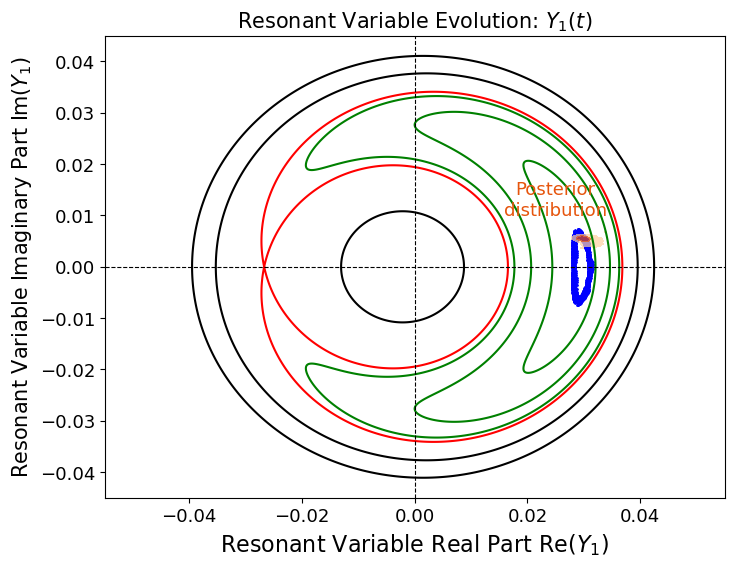}
\caption{Left panel: posterior distributions of the real and imaginary parts of $Y_1$ and $Y_2$ ($1\sigma$ and $2\sigma$ contours). $Y_1$ has much smaller uncertainties. Right panel: zoom in of left panel with the level curves of integrable approximate Hamiltonian described in \S\ref{sec:mmr}. The black and green contours show circulating and librating domains and the red line is the separatrix. The orange set represents the posterior distribution of $Y_1$ and blue points show the $N$-body integration described in \S\ref{sec:mmr} which confirm the system is in resonance with a small libration amplitude. \label{fig:Y1-Y2-post}}
\end{figure*}

\citet{2020MNRAS.496.3101P} plotted a rescaled version of this Hamiltonian which we reproduce in Figure~\ref{fig:Y1-Y2-post}. Here, $Y_1$ is a rescaled version of $\Psi_1$ and $\psi_1$ expressed as a complex variable:
\begin{equation}
Y_1 = \frac{\tilde{e}_b e^{i\varpi_b} - 1.22\,\tilde{e}_c e^{i\varpi_c}}{\sqrt{1 + 1.31\gamma}}\, e^{-i\theta_{\mathrm{res}}}
\label{eq:Y1}
\end{equation}
Here, $\gamma$ is the planet mass ratio $m_b/m_c$, $\theta_{\mathrm{res}}= 3 \lambda_c - 2 \lambda_b$, and $\tilde{e} = \sqrt{2\left(1 - \sqrt{1 - e^2}\right)}$.
\citet{2020MNRAS.496.3101P} defined an analogous $Y_2$ as 
\begin{equation}
Y_2 = \frac{1.07\gamma\,\tilde{e}_b e^{i\varpi_b} + \tilde{e}_c e^{i\varpi_c}}{\sqrt{1 + 1.31\gamma}}\, e^{-i\theta_{\mathrm{res}}}.
\label{eq:Y2}
\end{equation}

Since the dynamical evolution of two planets subject to Hamiltonian in Equation~\ref{eqn:hamiltonian-full} is described by $Y_1$ alone, we plot the Hamiltonian level sets in Figure~\ref{fig:Y1-Y2-post} along with draws from our credible models. We found that the planet resides in the librating portion of the phase-space diagram. The relevant resonant angle is

%

\begin{equation}
    \theta_1 
    = - \arg(Y_1) 
    = \theta_\mathrm{res} - \hat{\varpi},
\end{equation}
where
\begin{equation}
\hat{\varpi} = \arg(\tilde{e}_b e^{i\varpi_b} - 1.22\,\tilde{e}_c e^{i\varpi_c})
\end{equation}
and is sometimes called  the \emph{mixed pericenter angle} (see, e.g., \citealt{1986CeMec..38..175W}, \citealt{1986CeMec..38..335H}, \citealt{2013A&A...556A..28B}, \citealt{2023AJ....165...33D}). We confirmed that the system indeed librates by running a rebound simulation for $10^6$ planet b orbits. Its evolution is shown as a blue line in Figure~\ref{fig:Y1-Y2-post}, and we observe a libration amplitude of $\approx20$~deg.

Before wrapping up this section, we wish to pause and reflect on what dynamical quantities are constrained by the TTVs. The system evolves in the domain of $Y_1$ whereas $Y_2$ is a constant of motion. Since the TTVs, arise from this dynamical evolution, it is natural that they enable a tight constraint on $Y_1$ (see Figure~\ref{fig:Y1-Y2-post}). It is the small, but confidently non-zero value of $Y_1$, that rules out the possibility that both planets have zero eccentricity (see Figure~\ref{fig:ecc-post} and Equation~\ref{eq:Y1}).

From the perspective of a truly first-order model, we would not expect the TTVs to encode {\em any} information of $Y_2$. Thus, it is not surprising that the $Y_2$ is poorly constrained (see, again, Figure~\ref{fig:Y1-Y2-post}). We hypothesize that the loose constraints on $Y_2$ arise from higher order terms that are included in the $N$-body model. The connection between TTV observables and a system's resonant state is discussed further in \cite{2023ApJ...948...12G} and references therein.

\section{Formation and Evolution}
\label{sec:formation}

Mean motion resonance is a generic outcome of convergent migration. The high envelope mass fractions of roughly $15\%$ for planet c and $50\%$ for planet b \citep{2020AJ....159....2P} are consistent with formation in and migration through a gas-rich disk.

A notable puzzle, originally highlighted by \citet{2020MNRAS.496.3101P}, concerned the relatively high eccentricities reported by \citet{2020AJ....159....2P}. These eccentricities of around $\sim$0.2 would imply a significant angular momentum deficit (AMD) in the system. \citet{2020MNRAS.496.3101P} showed that AMD should be shared among the planets and transferred to the innermost planet, the Earth-size K2-19d on a 2.5~day orbit. This planet is close enough that one would expect eccentricity excitation to lead to tidal damping and orbital decay timescales shorter than the age of the system. 

\cite{2020MNRAS.496.3101P} defined the half-decay time, \(T_{\mathrm{hf}}\), as the duration over which the planet’s orbit decays by half of the total expected decay over 10 Gyr. \cite{2020MNRAS.496.3101P} found a median \(T_{\mathrm{hf}}\) of approximately 472 Myr, with 80\% of initial conditions exhibiting \(T_{\mathrm{hf}} < 1\) Gyr. This indicated that the majority of orbital decay occurs rapidly, on timescales much shorter than the system’s age. \cite{2020MNRAS.496.3101P} adopted a lower limit of 1~Gyr based on the rotation period and emerging gyrochronology relationships in 2020. These relationships are much better defined today, and we used the {\em gyro-interp} package \citep{2023ApJ...947L...3B} to derive an age of $2.4 \pm 0.3$~Gyr based on the measured rotation period and \teff. In any case, the presence of K2-19d more than a Gyr after formation was puzzling. 

\citet{2020MNRAS.496.3101P} hypothesized that the TTV model in \citet{2020AJ....159....2P} could be influenced by a potential non-transiting `planet e' between the orbits of planet d and b. We have not done a comprehensive search for such a planet in either the RVs or TTVs and cannot rule it out. However, a simpler resolution may lie in the structure of the solution space explored in previous models. Our updated dynamical analysis allows for configurations with significantly lower eccentricities for planets b and c, reducing the potential for angular momentum transfer to K2-19d. The low eccentricity range permitted by our models is insufficient to excite K2-19d toward tidal in-spiral, and may therefore help to explain the continued presence of this inner planet in the system. Indeed, \citet{2020MNRAS.496.3101P} estimated the minimum semi-major axis of planet d as a function of the initial AMD in the system and the planet current position (their Equation 15). We find that variation of more than 10\% of the orbital period of planet d by the low-eccentricity migration scenario requires planet b and c to have an eccentricity larger than 0.1, making the updated orbital configuration long-lived.

The low eccentricity is also consistent with the resonance being preserved over the lifetime of the system. Indeed, the tidal circularization timescale for planet b is 
\begin{equation}
    \tau_e = \frac{P_b}{21\pi}Q'_b\frac{m_b}{M_s}\left(\frac{a_b}{R_b}\right)^5 \label{eq:circu}
\end{equation}
\citep{Goldreich66} where $Q'=Q/k_2$ is the reduced tidal quality factor, $Q$ being the tidal quality factor and $k_2$ the Love number.
The Love number is usually unknown but of the order of a few  percent \citep{Batygin2025}.
Solving Equation~\ref{eq:circu} for $Q'$, imposing the circularization timescale to be longer than the system age ($\tau_e>T_\mathrm{sys} \simeq 2\ \mathrm{Gyr}$), we get a lower bound on $Q'\gtrsim 9.4\times10^3$.
This value is consistent with tidal quality factor for the gas and ice giants in the Solar System.

As tides tend to move planets away from the resonance \citep{Delisle2014}, the proximity of K2-19 b and c to the 3:2 MMR is noteworthy.
It indicates that the resonance state observed is likely pristine, with little evolution since the formation of the system and in particular since the disk dispersal. Indeed later on, only tides could provide a dissipative effect allowing to remain at low eccentricity and libration amplitude. 

Another noteworthy aspect is that the resonance persisted to the system's multi-Gyr age. \cite{2024AJ....168..239D} found the prevalence planets within a few percent of a first-order commensurability declines with stellar age: from $70 \pm 15$\% for young pairs (< 100~Myr) to $15 \pm 2$\% for mature pairs (>1 Gyr). The paucity of resonance in mature systems has motivated `breaking-the-chains' models where planets in resonance experience instabilities that lead to smooth distribution of period ratios observed among mature systems (e.g., \citealt{2017MNRAS.470.1750I}, \citealt{2020MNRAS.494.4950P}, \citealt{2022Icar..38815206G}; \citealt{2025AJ....169..323L}). If these instabilities are common outcomes of planetary evolution, the K2-19 system somehow escaped this fate. It may have been assisted by the unequal masses of planet b and c. Planets undergoing convergent migration may escape resonant locking due to overstable migration \citep{2014AJ....147...32G}  but resonant capture always remains stable if the inner planet is significantly more massive than the outer planet as in the case of K2-19 b \& c \citep{2015ApJ...810..119D,2025arXiv250112650L}.

\begin{figure}
    \centering
    \includegraphics[width=\linewidth]{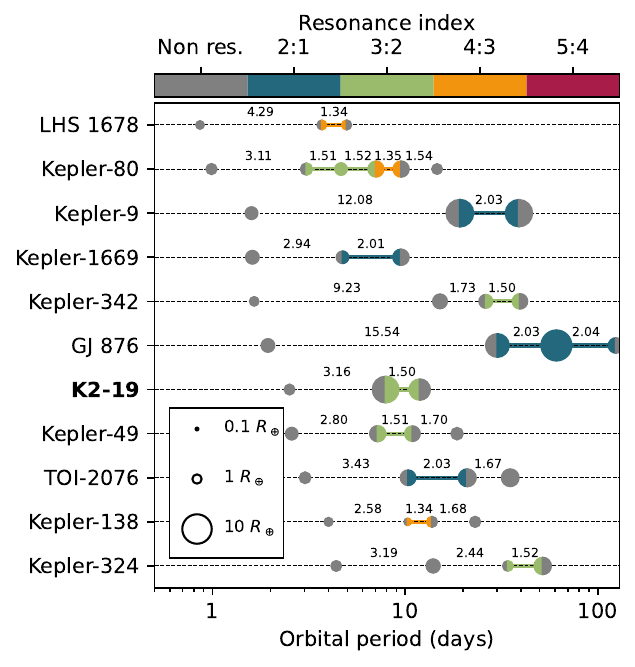}
    \caption{Architecture of systems composed of an inner planet detached from the rest of the system which comprises a resonant pair. The period ratio between adjacent planets is indicated over the line, with resonances highlighted in color.}
    \label{fig:orrery}
\end{figure}

The fact that the resonance coexists with the inner planet d without strong dynamical interactions gives strong constraints onto the system history. Indeed the large period ratio between d and b indicates that planet d could have been pushed by the arrival of the pair b and c at the inner edge of the disk \citep[as in the case of TRAPPIST-1,][]{Huang2022}. Such a configuration is not unique but rather rare, in particular for resonant Neptunes and gas giants.
We queried the Encyclopaedia of Exoplanetary Systems,%
\footnote{\href{http://exoplanet.eu/}{http://exoplanet.eu/}}
searching for systems composed of three or more planets, including an inner, detached planet (defined as having a period ratio larger than the 2:1 MMR and an orbital period smaller than 5 days) and a pair of planet in resonance. We plot the architectures of these systems in Figure~\ref{fig:orrery}, where the resonant index is indicated in color and the size of the planet is proportional to their radius.%
\footnote{When the radius is not known, we simply take the cubic root of the mass as this is mostly for visual illustration} 
The period ratio between adjacent planets is indicated over the line. Most of the systems selected from these criteria are composed of (super)-Earths only, K2-19 being one of the outliers, with Kepler-9 and GJ 876. Yet, those two other systems contains 2:1 MMR, which is more common among giant planets, highlighting the compactness of the K2-19 b-c pair.

\section{Conclusions}
\label{sec:conclusions}

We have presented an updated dynamical analysis of the K2-19 system, incorporating new transit-timing measurements from TESS along with previously published times from {\em K2}, {\em Spitzer}, and ground-based facilities. Using the gradient-based N-body integrator, {\em jnkepler}, we performed a comprehensive fit to the transit timing variations (TTVs) of planets K2-19b and c, enabling improved constraints on their orbital and physical properties. 

Our results confirm that K2-19b and K2-19c are engaged in a 3:2 mean-motion resonance and are actively librating within the resonant island. The proximity of the system to the nominal 3:2 resonance and the placement within the resonant phase portrait suggest that the planets are stably trapped, rather than near the separatrix or undergoing chaotic evolution. While the traditional resonant angles $\phi_1 = 3 \lambda_2 - 2 \lambda_1 - \varpi_{1}$ and $\phi_2 = 3 \lambda_2 - 2 \lambda_1 - \varpi_{2}$, do not librate, the libration of the argument of $Y_1$ provides a more robust diagnostic of resonance in this system. 

We found that either planets may have eccentricities ranging from $\approx$0.0--0.2 but both cannot have zero eccentricity. These constraints are broader than those reported in \cite{2020AJ....159....2P}. They partially alleviate previous concerns regarding the angular momentum transfer destabilizing the innermost planet, K2-19d, and help reconcile the system's long-term stability with its age of around 2-3~Gyr. Furthermore, the system's resonant configuration and moderate eccentricities are consistent with formation via convergent migration in a gas-rich protoplanetary disk. 

This work demonstrates the power of high-precision TTV analysis over extended baselines and highlights the importance of resonant variables in diagnosing dynamical states. Although we did not conduct a thorough search for a planet e, we note that its existence remains a possibility. Future transit detections of K2-19c, along with continued TTV monitoring and radial velocity follow-up, will be crucial in further probing the evolution of this compact, resonant planetary system. 
\\

We are grateful for conversations with Konstantin Batygin that improved this manuscript. Funding for this work was provided by a University of California, Los Angeles set-up award to E.A.P. and by the Heising-Simons Foundation Award \#2022-3833.

\facilities{TESS, Kepler}

\software 
{batman} \citep{2015PASP..127.1161K}, {jnkepler} \citep{2024AJ....168..294M}, {lightkurve} \citep{2018ascl.soft12013L}, {NumPyro} \citep{2018arXiv181009538B, 2019arXiv191211554P}, {NumPy} \citep{harris2020array}, {SciPy} \citep{2020SciPy-NMeth}, {Astropy} \citep{2013A&A...558A..33A, 2018AJ....156..123A, 2022ApJ...935..167A}.

\appendix
\section{Transit Times}
\label{sec:transit_times}
Table~\ref{tab:transit-times} lists our compilation of literature and {\em TESS} transit times.

\begin{deluxetable*}{lllrrrr}
\tablecaption{Transit Times\label{tab:transit-times}}
\tabletypesize{\footnotesize}
\tablehead{
  \colhead{Planet} & 
  \colhead{Transit} & 
  \colhead{Instrument} & 
  \colhead{$T_{c}$} & 
   \colhead{$\sigma(T_{c})$} &
  \colhead{Notes}   \\
  \colhead{} & 
  \colhead{} & 
  \colhead{} & 
  \colhead{days} & 
  \colhead{days} &
  \colhead{}
}
\startdata
b & -6 & K2 & 1980.3840 & 0.0004 & C \\
c & -3 & K2 & 1984.2723 & 0.0014 & C \\
b & -5 & K2 & 1988.3042 & 0.0006 & C \\
c & -2 & K2 & 1996.1841 & 0.0017 & C \\
b & -4 & K2 & 1996.2220 & 0.0005 & C \\
b & -3 & K2 & 2004.1385 & 0.0007 & C \\
c & -1 & K2 & 2008.0935 & 0.0020 & C \\
b & -2 & K2 & 2012.0616 & 0.0004 & C \\
b & -1 & K2 & 2019.9795 & 0.0007 & C \\
c & 0 & K2 & 2020.0022 & 0.0021 & C \\
b & 0 & K2 & 2027.9012 & 0.0007 & C \\
c & 1 & K2 & 2031.9071 & 0.0013 & C \\
b & 1 & K2 & 2035.8202 & 0.0004 & C \\
b & 2 & K2 & 2043.7393 & 0.0005 & C \\
c & 2 & K2 & 2043.8147 & 0.0014 & C \\
b & 3 & K2 & 2051.6586 & 0.0006 & C \\
c & 3 & K2 & 2055.7125 & 0.0019 & C \\
b & 24 & FLWO & 2218.0041 & 0.0022 & C \\
b & 28 & TRAPPIST & 2249.6955 & 0.0014 & C \\
b & 35 & MuSCAT & 2305.1505 & 0.0014 & C \\
c & 84 & Spitzer & 3019.4774 & 0.0074 & B \\
b & 127 & Spitzer & 3033.8604 & 0.0009 & B \\
b & 135 & LCO & 3097.2502 & 0.0024 & B \\
b & 144 & Spitzer & 3168.5368 & 0.0014 & B \\
c & 99 & Spitzer & 3197.8645 & 0.0059 & B \\
b & 337 & TESS & 4697.2890 & 0.0044 & A \\
b & 339 & TESS & 4713.1273 & 0.0028 & A \\
b & 340 & TESS & 4721.0418 & 0.0030 & A \\
b & 341 & TESS & 4728.9597 & 0.0030 & A \\
b & 342 & TESS & 4736.8808 & 0.0041 & A \\
b & 343 & TESS & 4744.7986 & 0.0028 & A \\
b & 430 & TESS & 5433.8788 & 0.0039 & A \\
b & 432 & TESS & 5449.7203 & 0.0023 & A \\

\enddata
\tablecomments{Following a convention from the \Kepler mission, times are given in $\mathrm{BJD}_\mathrm{TBD} - 2454833$. Notes---A: This work; B: \cite{2020AJ....159....2P}; C: \cite{2015ApJ...815...47N}}
\end{deluxetable*}

\section{Overlapping Transits}
While we did not attempt to extract transit times of planet c from {\em TESS} photometry, due to their low S/N, we checked for the possibility of simultaneous transits of b \& c that could affect our timing measurements of b. Figure~\ref{fig:c-prediction} shows that while three transits of planet b have nearby transits of c, none occur at the same time.

\label{sec:overlapping-transit}
\begin{figure*}
\centering
\includegraphics[width=1.0\textwidth]{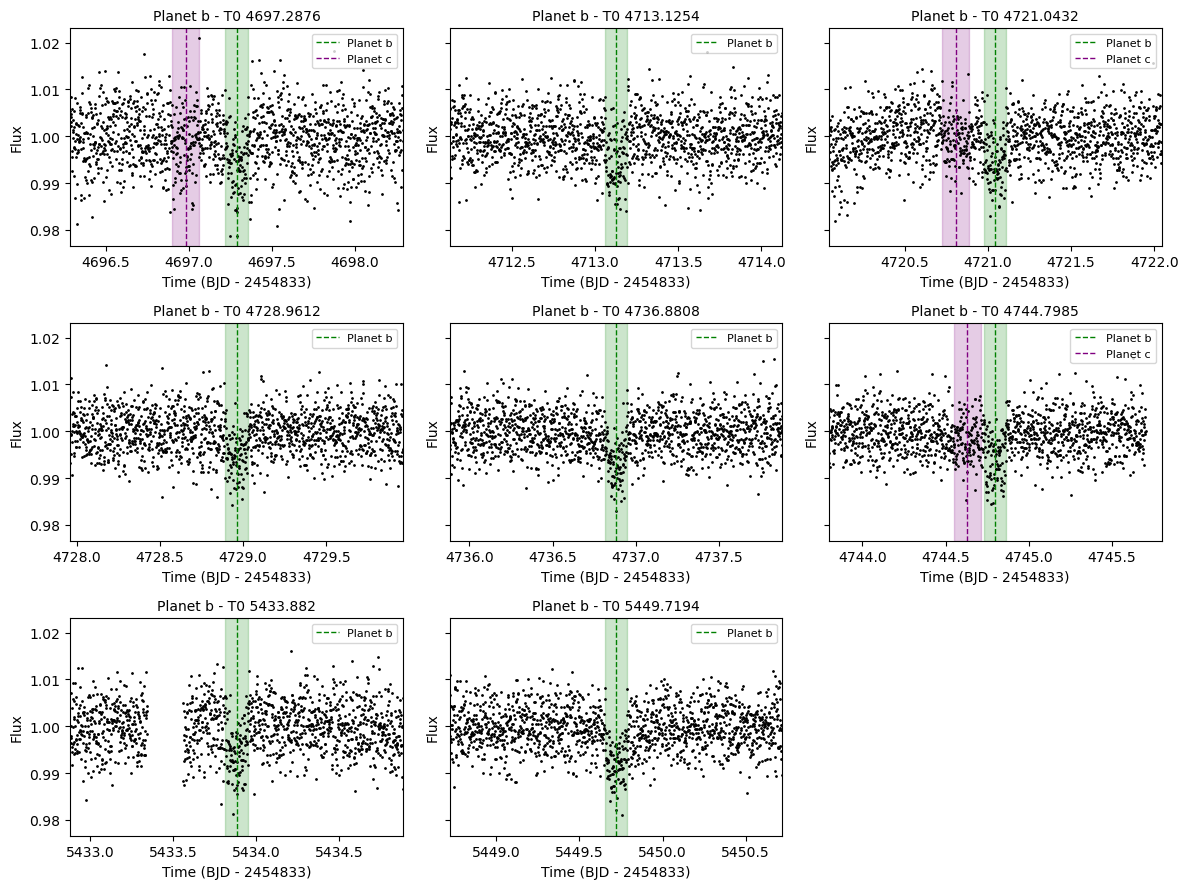}
\caption{Individual transit time predictions for K2-19b from the {\em jnkepler} model, with a window of 1~day before and after mid-transit. Predicted times for planet b (green) and planet c (purple) are shown by the dashed lines, and the shaded regions represent the respective transit durations. There are three planet b transit predictions in relative proximity to predictions for planet c transits. \label{fig:c-prediction}}
\end{figure*}

\section{Jnkepler Model Posteriors}
\label{sec:model-post}
\begin{figure*}
\centering
\includegraphics[width=1.0\textwidth]{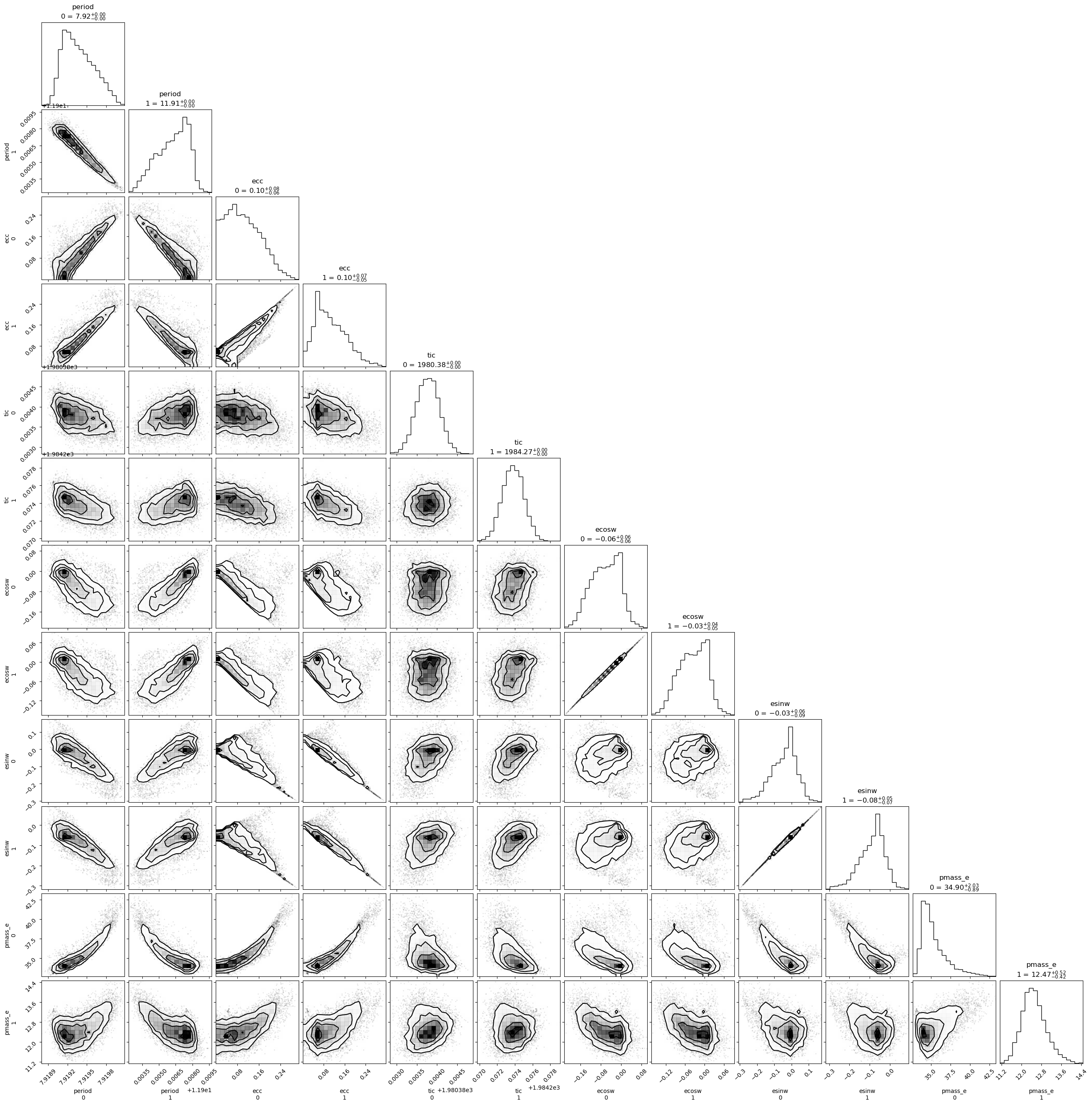}
\caption{2D joint posterior probability distributions for our TTV model (\S\ref{fig:jnkep-model}). \label{fig:model-post}}
\end{figure*}

Figure~\ref{fig:model-post} is a corner plot showing all the parameters in our TTV model.

\bibliography{manuscript.bib}

\begin{thebibliography}{}
\expandafter\ifx\csname natexlab\endcsname\relax\def\natexlab#1{#1}\fi
\providecommand{\url}[1]{\href{#1}{#1}}
\providecommand{\dodoi}[1]{doi:~\href{http://doi.org/#1}{\nolinkurl{#1}}}
\providecommand{\doeprint}[1]{\href{http://ascl.net/#1}{\nolinkurl{http://ascl.net/#1}}}
\providecommand{\doarXiv}[1]{\href{https://arxiv.org/abs/#1}{\nolinkurl{https://arxiv.org/abs/#1}}}

\bibitem[{{Akeson} {et~al.}(2013){Akeson}, {Chen}, {Ciardi}, {Crane}, {Good}, {Harbut}, {Jackson}, {Kane}, {Laity}, {Leifer}, {Lynn}, {McElroy}, {Papin}, {Plavchan}, {Ram{\'\i}rez}, {Rey}, {von Braun}, {Wittman}, {Abajian}, {Ali}, {Beichman}, {Beekley}, {Berriman}, {Berukoff}, {Bryden}, {Chan}, {Groom}, {Lau}, {Payne}, {Regelson}, {Saucedo}, {Schmitz}, {Stauffer}, {Wyatt}, \& {Zhang}}]{2013PASP..125..989A}
{Akeson}, R.~L., {Chen}, X., {Ciardi}, D., {et~al.} 2013, \pasp, 125, 989, \dodoi{10.1086/672273}

\bibitem[{{Armstrong} {et~al.}(2015){Armstrong}, {Santerne}, {Veras}, {Barros}, {Demangeon}, {Lillo-Box}, {McCormac}, {Osborn}, {Tsantaki}, {Almenara}, {Barrado}, {Boisse}, {Bonomo}, {Brown}, {Bruno}, {Rey Cerda}, {Courcol}, {Deleuil}, {D{\'\i}az}, {Doyle}, {H{\'e}brard}, {Kirk}, {Lam}, {Pollacco}, {Rajpurohit}, {Spake}, \& {Walker}}]{2015A&A...582A..33A}
{Armstrong}, D.~J., {Santerne}, A., {Veras}, D., {et~al.} 2015, \aap, 582, A33, \dodoi{10.1051/0004-6361/201526008}

\bibitem[{{Astropy Collaboration} {et~al.}(2013){Astropy Collaboration}, {Robitaille}, {Tollerud}, {Greenfield}, {Droettboom}, {Bray}, {Aldcroft}, {Davis}, {Ginsburg}, {Price-Whelan}, {Kerzendorf}, {Conley}, {Crighton}, {Barbary}, {Muna}, {Ferguson}, {Grollier}, {Parikh}, {Nair}, {Unther}, {Deil}, {Woillez}, {Conseil}, {Kramer}, {Turner}, {Singer}, {Fox}, {Weaver}, {Zabalza}, {Edwards}, {Azalee Bostroem}, {Burke}, {Casey}, {Crawford}, {Dencheva}, {Ely}, {Jenness}, {Labrie}, {Lim}, {Pierfederici}, {Pontzen}, {Ptak}, {Refsdal}, {Servillat}, \& {Streicher}}]{2013A&A...558A..33A}
{Astropy Collaboration}, {Robitaille}, T.~P., {Tollerud}, E.~J., {et~al.} 2013, \aap, 558, A33, \dodoi{10.1051/0004-6361/201322068}

\bibitem[{{Astropy Collaboration} {et~al.}(2018){Astropy Collaboration}, {Price-Whelan}, {Sip{\H{o}}cz}, {G{\"u}nther}, {Lim}, {Crawford}, {Conseil}, {Shupe}, {Craig}, {Dencheva}, {Ginsburg}, {VanderPlas}, {Bradley}, {P{\'e}rez-Su{\'a}rez}, {de Val-Borro}, {Aldcroft}, {Cruz}, {Robitaille}, {Tollerud}, {Ardelean}, {Babej}, {Bach}, {Bachetti}, {Bakanov}, {Bamford}, {Barentsen}, {Barmby}, {Baumbach}, {Berry}, {Biscani}, {Boquien}, {Bostroem}, {Bouma}, {Brammer}, {Bray}, {Breytenbach}, {Buddelmeijer}, {Burke}, {Calderone}, {Cano Rodr{\'\i}guez}, {Cara}, {Cardoso}, {Cheedella}, {Copin}, {Corrales}, {Crichton}, {D'Avella}, {Deil}, {Depagne}, {Dietrich}, {Donath}, {Droettboom}, {Earl}, {Erben}, {Fabbro}, {Ferreira}, {Finethy}, {Fox}, {Garrison}, {Gibbons}, {Goldstein}, {Gommers}, {Greco}, {Greenfield}, {Groener}, {Grollier}, {Hagen}, {Hirst}, {Homeier}, {Horton}, {Hosseinzadeh}, {Hu}, {Hunkeler}, {Ivezi{\'c}}, {Jain}, {Jenness}, {Kanarek}, {Kendrew}, {Kern}, {Kerzendorf}, {Khvalko}, {King}, {Kirkby}, {Kulkarni},
  {Kumar}, {Lee}, {Lenz}, {Littlefair}, {Ma}, {Macleod}, {Mastropietro}, {McCully}, {Montagnac}, {Morris}, {Mueller}, {Mumford}, {Muna}, {Murphy}, {Nelson}, {Nguyen}, {Ninan}, {N{\"o}the}, {Ogaz}, {Oh}, {Parejko}, {Parley}, {Pascual}, {Patil}, {Patil}, {Plunkett}, {Prochaska}, {Rastogi}, {Reddy Janga}, {Sabater}, {Sakurikar}, {Seifert}, {Sherbert}, {Sherwood-Taylor}, {Shih}, {Sick}, {Silbiger}, {Singanamalla}, {Singer}, {Sladen}, {Sooley}, {Sornarajah}, {Streicher}, {Teuben}, {Thomas}, {Tremblay}, {Turner}, {Terr{\'o}n}, {van Kerkwijk}, {de la Vega}, {Watkins}, {Weaver}, {Whitmore}, {Woillez}, {Zabalza}, \& {Astropy Contributors}}]{2018AJ....156..123A}
{Astropy Collaboration}, {Price-Whelan}, A.~M., {Sip{\H{o}}cz}, B.~M., {et~al.} 2018, \aj, 156, 123, \dodoi{10.3847/1538-3881/aabc4f}

\bibitem[{{Astropy Collaboration} {et~al.}(2022){Astropy Collaboration}, {Price-Whelan}, {Lim}, {Earl}, {Starkman}, {Bradley}, {Shupe}, {Patil}, {Corrales}, {Brasseur}, {N{\"o}the}, {Donath}, {Tollerud}, {Morris}, {Ginsburg}, {Vaher}, {Weaver}, {Tocknell}, {Jamieson}, {van Kerkwijk}, {Robitaille}, {Merry}, {Bachetti}, {G{\"u}nther}, {Aldcroft}, {Alvarado-Montes}, {Archibald}, {B{\'o}di}, {Bapat}, {Barentsen}, {Baz{\'a}n}, {Biswas}, {Boquien}, {Burke}, {Cara}, {Cara}, {Conroy}, {Conseil}, {Craig}, {Cross}, {Cruz}, {D'Eugenio}, {Dencheva}, {Devillepoix}, {Dietrich}, {Eigenbrot}, {Erben}, {Ferreira}, {Foreman-Mackey}, {Fox}, {Freij}, {Garg}, {Geda}, {Glattly}, {Gondhalekar}, {Gordon}, {Grant}, {Greenfield}, {Groener}, {Guest}, {Gurovich}, {Handberg}, {Hart}, {Hatfield-Dodds}, {Homeier}, {Hosseinzadeh}, {Jenness}, {Jones}, {Joseph}, {Kalmbach}, {Karamehmetoglu}, {Ka{\l}uszy{\'n}ski}, {Kelley}, {Kern}, {Kerzendorf}, {Koch}, {Kulumani}, {Lee}, {Ly}, {Ma}, {MacBride}, {Maljaars}, {Muna}, {Murphy}, {Norman},
  {O'Steen}, {Oman}, {Pacifici}, {Pascual}, {Pascual-Granado}, {Patil}, {Perren}, {Pickering}, {Rastogi}, {Roulston}, {Ryan}, {Rykoff}, {Sabater}, {Sakurikar}, {Salgado}, {Sanghi}, {Saunders}, {Savchenko}, {Schwardt}, {Seifert-Eckert}, {Shih}, {Jain}, {Shukla}, {Sick}, {Simpson}, {Singanamalla}, {Singer}, {Singhal}, {Sinha}, {Sip{\H{o}}cz}, {Spitler}, {Stansby}, {Streicher}, {{\v{S}}umak}, {Swinbank}, {Taranu}, {Tewary}, {Tremblay}, {de Val-Borro}, {Van Kooten}, {Vasovi{\'c}}, {Verma}, {de Miranda Cardoso}, {Williams}, {Wilson}, {Winkel}, {Wood-Vasey}, {Xue}, {Yoachim}, {Zhang}, {Zonca}, \& {Astropy Project Contributors}}]{2022ApJ...935..167A}
{Astropy Collaboration}, {Price-Whelan}, A.~M., {Lim}, P.~L., {et~al.} 2022, \apj, 935, 167, \dodoi{10.3847/1538-4357/ac7c74}

\bibitem[{{Barros} {et~al.}(2015){Barros}, {Almenara}, {Demangeon}, {Tsantaki}, {Santerne}, {Armstrong}, {Barrado}, {Brown}, {Deleuil}, {Lillo-Box}, {Osborn}, {Pollacco}, {Abe}, {Andre}, {Bendjoya}, {Boisse}, {Bonomo}, {Bouchy}, {Bruno}, {Cerda}, {Courcol}, {D{\'\i}az}, {H{\'e}brard}, {Kirk}, {Lachuri{\'e}}, {Lam}, {Martinez}, {McCormac}, {Moutou}, {Rajpurohit}, {Rivet}, {Spake}, {Suarez}, {Toublanc}, \& {Walker}}]{2015MNRAS.454.4267B}
{Barros}, S.~C.~C., {Almenara}, J.~M., {Demangeon}, O., {et~al.} 2015, \mnras, 454, 4267, \dodoi{10.1093/mnras/stv2271}

\bibitem[{Batygin(2025)}]{Batygin2025}
Batygin, K. 2025, The Astrophysical Journal, 985, 87, \dodoi{10.3847/1538-4357/adccc4}

\bibitem[{{Batygin} \& {Morbidelli}(2013)}]{2013A&A...556A..28B}
{Batygin}, K., \& {Morbidelli}, A. 2013, \aap, 556, A28, \dodoi{10.1051/0004-6361/201220907}

\bibitem[{{Betancourt}(2017)}]{2017arXiv170102434B}
{Betancourt}, M. 2017, arXiv e-prints, arXiv:1701.02434, \dodoi{10.48550/arXiv.1701.02434}

\bibitem[{{Bingham} {et~al.}(2018){Bingham}, {Chen}, {Jankowiak}, {Obermeyer}, {Pradhan}, {Karaletsos}, {Singh}, {Szerlip}, {Horsfall}, \& {Goodman}}]{2018arXiv181009538B}
{Bingham}, E., {Chen}, J.~P., {Jankowiak}, M., {et~al.} 2018, arXiv e-prints, arXiv:1810.09538, \dodoi{10.48550/arXiv.1810.09538}

\bibitem[{{Bouma} {et~al.}(2023){Bouma}, {Palumbo}, \& {Hillenbrand}}]{2023ApJ...947L...3B}
{Bouma}, L.~G., {Palumbo}, E.~K., \& {Hillenbrand}, L.~A. 2023, \apjl, 947, L3, \dodoi{10.3847/2041-8213/acc589}

\bibitem[{Bradbury {et~al.}(2018)Bradbury, Frostig, Hawkins, Johnson, Leary, Maclaurin, Necula, Paszke, Vander{P}las, Wanderman-{M}ilne, \& Zhang}]{jax2018github}
Bradbury, J., Frostig, R., Hawkins, P., {et~al.} 2018, {JAX}: composable transformations of {P}ython+{N}um{P}y programs, 0.3.13.
\newblock \url{http://github.com/jax-ml/jax}

\bibitem[{{Dai} {et~al.}(2023){Dai}, {Masuda}, {Beard}, {Robertson}, {Goldberg}, {Batygin}, {Bouma}, {Lissauer}, {Knudstrup}, {Albrecht}, {Howard}, {Knutson}, {Petigura}, {Weiss}, {Isaacson}, {Kristiansen}, {Osborn}, {Wang}, {Wang}, {Behmard}, {Greklek-McKeon}, {Vissapragada}, {Batalha}, {Brinkman}, {Chontos}, {Crossfield}, {Dressing}, {Fetherolf}, {Fulton}, {Hill}, {Huber}, {Kane}, {Lubin}, {MacDougall}, {Mayo}, {Mo{\v{c}}nik}, {Akana Murphy}, {Rubenzahl}, {Scarsdale}, {Tyler}, {Zandt}, {Polanski}, {Schwengeler}, {Terentev}, {Benni}, {Bieryla}, {Ciardi}, {Falk}, {Furlan}, {Girardin}, {Guerra}, {Hesse}, {Howell}, {Lillo-Box}, {Matthews}, {Twicken}, {Villase{\~n}or}, {Latham}, {Jenkins}, {Ricker}, {Seager}, {Vanderspek}, \& {Winn}}]{2023AJ....165...33D}
{Dai}, F., {Masuda}, K., {Beard}, C., {et~al.} 2023, \aj, 165, 33, \dodoi{10.3847/1538-3881/aca327}

\bibitem[{{Dai} {et~al.}(2024){Dai}, {Goldberg}, {Batygin}, {van Saders}, {Chiang}, {Choksi}, {Li}, {Petigura}, {Gilbert}, {Millholland}, {Dai}, {Bouma}, {Weiss}, \& {Winn}}]{2024AJ....168..239D}
{Dai}, F., {Goldberg}, M., {Batygin}, K., {et~al.} 2024, \aj, 168, 239, \dodoi{10.3847/1538-3881/ad83a6}

\bibitem[{{Deck} \& {Batygin}(2015)}]{2015ApJ...810..119D}
{Deck}, K.~M., \& {Batygin}, K. 2015, \apj, 810, 119, \dodoi{10.1088/0004-637X/810/2/119}

\bibitem[{Delisle {et~al.}(2014)Delisle, Laskar, \& Correia}]{Delisle2014}
Delisle, J.-B., Laskar, J., \& Correia, A. C.~M. 2014, Astronomy \& Astrophysics, 566, A137, \dodoi{10.1051/0004-6361/201423676}

\bibitem[{{Duane} {et~al.}(1987){Duane}, {Kennedy}, {Pendleton}, \& {Roweth}}]{1987PhLB..195..216D}
{Duane}, S., {Kennedy}, A.~D., {Pendleton}, B.~J., \& {Roweth}, D. 1987, Physics Letters B, 195, 216, \dodoi{10.1016/0370-2693(87)91197-X}

\bibitem[{{Gelman} \& {Rubin}(1992)}]{1992StaSc...7..457G}
{Gelman}, A., \& {Rubin}, D.~B. 1992, Statistical Science, 7, 457, \dodoi{10.1214/ss/1177011136}

\bibitem[{{Goldberg} \& {Batygin}(2023)}]{2023ApJ...948...12G}
{Goldberg}, M., \& {Batygin}, K. 2023, \apj, 948, 12, \dodoi{10.3847/1538-4357/acc9ae}

\bibitem[{{Goldberg} {et~al.}(2022){Goldberg}, {Batygin}, \& {Morbidelli}}]{2022Icar..38815206G}
{Goldberg}, M., {Batygin}, K., \& {Morbidelli}, A. 2022, \icarus, 388, 115206, \dodoi{10.1016/j.icarus.2022.115206}

\bibitem[{{Goldreich} \& {Schlichting}(2014)}]{2014AJ....147...32G}
{Goldreich}, P., \& {Schlichting}, H.~E. 2014, \aj, 147, 32, \dodoi{10.1088/0004-6256/147/2/32}

\bibitem[{{Goldreich} \& {Soter}(1966)}]{Goldreich66}
{Goldreich}, P., \& {Soter}, S. 1966, \icarus, 5, 375, \dodoi{10.1016/0019-1035(66)90051-0}

\bibitem[{Harris {et~al.}(2020)Harris, Millman, van~der Walt, Gommers, Virtanen, Cournapeau, Wieser, Taylor, Berg, Smith, Kern, Picus, Hoyer, van Kerkwijk, Brett, Haldane, del R{\'{i}}o, Wiebe, Peterson, G{\'{e}}rard-Marchant, Sheppard, Reddy, Weckesser, Abbasi, Gohlke, \& Oliphant}]{harris2020array}
Harris, C.~R., Millman, K.~J., van~der Walt, S.~J., {et~al.} 2020, Nature, 585, 357, \dodoi{10.1038/s41586-020-2649-2}

\bibitem[{{Henrard} {et~al.}(1986){Henrard}, {Lemaitre}, {Milani}, \& {Murray}}]{1986CeMec..38..335H}
{Henrard}, J., {Lemaitre}, A., {Milani}, A., \& {Murray}, C.~D. 1986, Celestial Mechanics, 38, 335, \dodoi{10.1007/BF01238924}

\bibitem[{{Hoffman} \& {Gelman}(2011)}]{2011arXiv1111.4246H}
{Hoffman}, M.~D., \& {Gelman}, A. 2011, arXiv e-prints, arXiv:1111.4246, \dodoi{10.48550/arXiv.1111.4246}

\bibitem[{{Holczer} {et~al.}(2016){Holczer}, {Mazeh}, {Nachmani}, {Jontof-Hutter}, {Ford}, {Fabrycky}, {Ragozzine}, {Kane}, \& {Steffen}}]{2016ApJS..225....9H}
{Holczer}, T., {Mazeh}, T., {Nachmani}, G., {et~al.} 2016, \apjs, 225, 9, \dodoi{10.3847/0067-0049/225/1/9}

\bibitem[{Huang \& Ormel(2022)}]{Huang2022}
Huang, S., \& Ormel, C.~W. 2022, Monthly Notices of the Royal Astronomical Society, 511, 3814, \dodoi{10.1093/mnras/stac288}

\bibitem[{Husser {et~al.}(2013)Husser, {Wende-von Berg}, Dreizler, Homeier, Reiners, Barman, \& Hauschildt}]{Husser2013}
Husser, T.-O., {Wende-von Berg}, S., Dreizler, S., {et~al.} 2013, A{\&}A, 553, A6, \dodoi{10.1051/0004-6361/201219058}

\bibitem[{{Izidoro} {et~al.}(2017){Izidoro}, {Ogihara}, {Raymond}, {Morbidelli}, {Pierens}, {Bitsch}, {Cossou}, \& {Hersant}}]{2017MNRAS.470.1750I}
{Izidoro}, A., {Ogihara}, M., {Raymond}, S.~N., {et~al.} 2017, \mnras, 470, 1750, \dodoi{10.1093/mnras/stx1232}

\bibitem[{{Kreidberg}(2015)}]{2015PASP..127.1161K}
{Kreidberg}, L. 2015, \pasp, 127, 1161, \dodoi{10.1086/683602}

\bibitem[{{Lammers} \& {Winn}(2024)}]{2024ApJ...968L..12L}
{Lammers}, C., \& {Winn}, J.~N. 2024, \apjl, 968, L12, \dodoi{10.3847/2041-8213/ad50d2}

\bibitem[{{Laplace} {et~al.}(1829){Laplace}, {Bowditch}, \& {Bowditch}}]{1829mecc.book.....L}
{Laplace}, P.~S., {Bowditch}, N., \& {Bowditch}, N.~I. 1829, {M{\'e}canique c{\'e}leste}

\bibitem[{{Laskar}(1997)}]{1997A&A...317L..75L}
{Laskar}, J. 1997, \aap, 317, L75

\bibitem[{{Lee} \& {Peale}(2001)}]{2001astro.ph..8104L}
{Lee}, M.~H., \& {Peale}, S.~J. 2001, arXiv e-prints, astro, \dodoi{10.48550/arXiv.astro-ph/0108104}

\bibitem[{{Li} {et~al.}(2025){Li}, {Chiang}, {Choksi}, \& {Dai}}]{2025AJ....169..323L}
{Li}, R., {Chiang}, E., {Choksi}, N., \& {Dai}, F. 2025, \aj, 169, 323, \dodoi{10.3847/1538-3881/adce0c}

\bibitem[{{Lightkurve Collaboration} {et~al.}(2018){Lightkurve Collaboration}, {Cardoso}, {Hedges}, {Gully-Santiago}, {Saunders}, {Cody}, {Barclay}, {Hall}, {Sagear}, {Turtelboom}, {Zhang}, {Tzanidakis}, {Mighell}, {Coughlin}, {Bell}, {Berta-Thompson}, {Williams}, {Dotson}, \& {Barentsen}}]{2018ascl.soft12013L}
{Lightkurve Collaboration}, {Cardoso}, J.~V.~d.~M., {Hedges}, C., {et~al.} 2018, {Lightkurve: Kepler and TESS time series analysis in Python}, Astrophysics Source Code Library.
\newblock \doeprint{1812.013}

\bibitem[{{Lin} {et~al.}(2025){Lin}, {Liu}, \& {Zheng}}]{2025arXiv250112650L}
{Lin}, L., {Liu}, B., \& {Zheng}, Z. 2025, arXiv e-prints, arXiv:2501.12650, \dodoi{10.48550/arXiv.2501.12650}

\bibitem[{{Luger} {et~al.}(2017){Luger}, {Sestovic}, {Kruse}, {Grimm}, {Demory}, {Agol}, {Bolmont}, {Fabrycky}, {Fernandes}, {Van Grootel}, {Burgasser}, {Gillon}, {Ingalls}, {Jehin}, {Raymond}, {Selsis}, {Triaud}, {Barclay}, {Barentsen}, {Howell}, {Delrez}, {de Wit}, {Foreman-Mackey}, {Holdsworth}, {Leconte}, {Lederer}, {Turbet}, {Almleaky}, {Benkhaldoun}, {Magain}, {Morris}, {Heng}, \& {Queloz}}]{2017NatAs...1E.129L}
{Luger}, R., {Sestovic}, M., {Kruse}, E., {et~al.} 2017, Nature Astronomy, 1, 0129, \dodoi{10.1038/s41550-017-0129}

\bibitem[{{Mandel} \& {Agol}(2002)}]{2002ApJ...580L.171M}
{Mandel}, K., \& {Agol}, E. 2002, \apjl, 580, L171, \dodoi{10.1086/345520}

\bibitem[{{Masuda} {et~al.}(2024){Masuda}, {Libby-Roberts}, {Livingston}, {Stevenson}, {Gao}, {Vissapragada}, {Fu}, {Han}, {Greklek-McKeon}, {Mahadevan}, {Agol}, {Bello-Arufe}, {Berta-Thompson}, {Ca{\~n}as}, {Chachan}, {Hebb}, {Hu}, {Kawashima}, {Knutson}, {Morley}, {Murray}, {Ohno}, {Tokadjian}, {Zhang}, {Welbanks}, {Nixon}, {Freedman}, {Narita}, {Fukui}, {de Leon}, {Mori}, {Palle}, {Murgas}, {Parviainen}, {Esparza-Borges}, {Jontof-Hutter}, {Collins}, {Benni}, {Barkaoui}, {Pozuelos}, {Gillon}, {Jehin}, {Benkhaldoun}, {Hawley}, {Lin}, {Stef{\'a}nsson}, {Bieryla}, {Yilmaz}, {Senavci}, {Girardin}, {Marino}, \& {Wang}}]{2024AJ....168..294M}
{Masuda}, K., {Libby-Roberts}, J.~E., {Livingston}, J.~H., {et~al.} 2024, \aj, 168, 294, \dodoi{10.3847/1538-3881/ad83d3}

\bibitem[{{Mazeh} {et~al.}(2013){Mazeh}, {Nachmani}, {Holczer}, {Fabrycky}, {Ford}, {Sanchis-Ojeda}, {Sokol}, {Rowe}, {Zucker}, {Agol}, {Carter}, {Lissauer}, {Quintana}, {Ragozzine}, {Steffen}, \& {Welsh}}]{2013ApJS..208...16M}
{Mazeh}, T., {Nachmani}, G., {Holczer}, T., {et~al.} 2013, \apjs, 208, 16, \dodoi{10.1088/0067-0049/208/2/16}

\bibitem[{{Mills} {et~al.}(2016){Mills}, {Fabrycky}, {Migaszewski}, {Ford}, {Petigura}, \& {Isaacson}}]{2016Natur.533..509M}
{Mills}, S.~M., {Fabrycky}, D.~C., {Migaszewski}, C., {et~al.} 2016, \nat, 533, 509, \dodoi{10.1038/nature17445}

\bibitem[{{Narita} {et~al.}(2015){Narita}, {Hirano}, {Fukui}, {Hori}, {Sanchis-Ojeda}, {Winn}, {Ryu}, {Kusakabe}, {Kudo}, {Onitsuka}, {Delrez}, {Gillon}, {Jehin}, {McCormac}, {Holman}, {Izumiura}, {Takeda}, {Tamura}, \& {Yanagisawa}}]{2015ApJ...815...47N}
{Narita}, N., {Hirano}, T., {Fukui}, A., {et~al.} 2015, \apj, 815, 47, \dodoi{10.1088/0004-637X/815/1/47}

\bibitem[{{Nesvorn{\'y}} \& {Vokrouhlick{\'y}}(2016)}]{2016ApJ...823...72N}
{Nesvorn{\'y}}, D., \& {Vokrouhlick{\'y}}, D. 2016, \apj, 823, 72, \dodoi{10.3847/0004-637X/823/2/72}

\bibitem[{Parviainen \& Aigrain(2015)}]{Parviainen2015}
Parviainen, H., \& Aigrain, S. 2015, MNRAS, 453, 3821, \dodoi{10.1093/mnras/stv1857}

\bibitem[{{Petigura} {et~al.}(2018){Petigura}, {Marcy}, {Winn}, {Weiss}, {Fulton}, {Howard}, {Sinukoff}, {Isaacson}, {Morton}, \& {Johnson}}]{2018AJ....155...89P}
{Petigura}, E.~A., {Marcy}, G.~W., {Winn}, J.~N., {et~al.} 2018, \aj, 155, 89, \dodoi{10.3847/1538-3881/aaa54c}

\bibitem[{{Petigura} {et~al.}(2020){Petigura}, {Livingston}, {Batygin}, {Mills}, {Werner}, {Isaacson}, {Fulton}, {Howard}, {Weiss}, {Espinoza}, {Jontof-Hutter}, {Shporer}, {Bayliss}, \& {Barros}}]{2020AJ....159....2P}
{Petigura}, E.~A., {Livingston}, J., {Batygin}, K., {et~al.} 2020, \aj, 159, 2, \dodoi{10.3847/1538-3881/ab5220}

\bibitem[{{Petit} {et~al.}(2020){Petit}, {Petigura}, {Davies}, \& {Johansen}}]{2020MNRAS.496.3101P}
{Petit}, A.~C., {Petigura}, E.~A., {Davies}, M.~B., \& {Johansen}, A. 2020, \mnras, 496, 3101, \dodoi{10.1093/mnras/staa1736}

\bibitem[{{Phan} {et~al.}(2019){Phan}, {Pradhan}, \& {Jankowiak}}]{2019arXiv191211554P}
{Phan}, D., {Pradhan}, N., \& {Jankowiak}, M. 2019, arXiv e-prints, arXiv:1912.11554, \dodoi{10.48550/arXiv.1912.11554}

\bibitem[{{Pichierri} \& {Morbidelli}(2020)}]{2020MNRAS.494.4950P}
{Pichierri}, G., \& {Morbidelli}, A. 2020, \mnras, 494, 4950, \dodoi{10.1093/mnras/staa1102}

\bibitem[{{Press} {et~al.}(1992){Press}, {Teukolsky}, {Vetterling}, \& {Flannery}}]{1992nrfa.book.....P}
{Press}, W.~H., {Teukolsky}, S.~A., {Vetterling}, W.~T., \& {Flannery}, B.~P. 1992, {Numerical recipes in FORTRAN. The art of scientific computing}

\bibitem[{{Sessin} \& {Ferraz-Mello}(1984)}]{1984CeMec..32..307S}
{Sessin}, W., \& {Ferraz-Mello}, S. 1984, Celestial Mechanics, 32, 307, \dodoi{10.1007/BF01229087}

\bibitem[{{Sinukoff} {et~al.}(2016){Sinukoff}, {Howard}, {Petigura}, {Schlieder}, {Crossfield}, {Ciardi}, {Fulton}, {Isaacson}, {Aller}, {Baranec}, {Beichman}, {Hansen}, {Knutson}, {Law}, {Liu}, {Riddle}, \& {Dressing}}]{2016ApJ...827...78S}
{Sinukoff}, E., {Howard}, A.~W., {Petigura}, E.~A., {et~al.} 2016, \apj, 827, 78, \dodoi{10.3847/0004-637X/827/1/78}

\bibitem[{Virtanen {et~al.}(2020)Virtanen, Gommers, Oliphant, Haberland, Reddy, Cournapeau, Burovski, Peterson, Weckesser, Bright, {van der Walt}, Brett, Wilson, Millman, Mayorov, Nelson, Jones, Kern, Larson, Carey, Polat, Feng, Moore, {VanderPlas}, Laxalde, Perktold, Cimrman, Henriksen, Quintero, Harris, Archibald, Ribeiro, Pedregosa, {van Mulbregt}, \& {SciPy 1.0 Contributors}}]{2020SciPy-NMeth}
Virtanen, P., Gommers, R., Oliphant, T.~E., {et~al.} 2020, Nature Methods, 17, 261, \dodoi{10.1038/s41592-019-0686-2}

\bibitem[{{Wisdom}(1986)}]{1986CeMec..38..175W}
{Wisdom}, J. 1986, Celestial Mechanics, 38, 175, \dodoi{10.1007/BF01230429}

\bibitem[{{Wisdom} \& {Holman}(1991)}]{1991AJ....102.1528W}
{Wisdom}, J., \& {Holman}, M. 1991, \aj, 102, 1528, \dodoi{10.1086/115978}

\end{thebibliography}

\end{document}